\documentclass[aps,prl,showpacs,twocolumn,reprint,floats,epsfig,pdflatex]{revtex4-1}
\usepackage{epsfig}
\usepackage{amsmath}
\usepackage{amsfonts}
\usepackage{graphicx}
\usepackage{amssymb}
\usepackage{amsbsy}
\usepackage{subfigure}

\newcommand{\ket}[1]{|{#1}\rangle}

\begin{document}

\title {Ramp dynamics of phonons in an ion trap: entanglement generation and cooling}

\author{T. Dutta$^{(1)}$, M. Mukherjee$^{(1,2)}$, and K. Sengupta$^{(3)}$}
\affiliation{$^{(1)}$ Centre for Quantum Technologies, National
University Singapore, Singapore 117543, Singapore. \\ $^{(2)}$
Physics Department, National University of Singapore, 2 Science
Drive 3 Singapore 117551, Singapore. \\ $^{(3)}$ Theoretical Physics
Department, Indian Association for the Cultivation of Science,
Jadavpur, Kolkata-700032, India.}

\date{\today}

\begin{abstract}

We show that the ramp dynamics of phonons in an one-dimensional ion
trap can be used for both generating multi-particle entangled states
and motional state cooling of a string of trapped ions. We study
such ramp dynamics using an effective Bose-Hubbard model which
describes these phonons at low energies and show that specific
protocols, involving site-specific dynamical tuning of the on-site
potential of the model, can be used to generate entangled states and
to achieve motional state cooling without involving electronic
states of the ions. We compare and contrast our schemes for these to
the earlier suggested ones and discuss specific experiments to
realize the suggested protocols.

\end{abstract}

\pacs{03.75.Lm, 05.30.Jp, 05.30.Rt}

\maketitle

Emulation of isolated strongly correlated quantum systems has been a
subject of intense experimental and theoretical research in the
recent past \cite{bloch1,zoller1,cirac1,monroe1, greiner1,sengupta1,
bakr1}. The two most easily realizable models discussed in this
context are the Ising and the Bose-Hubbard model (BHM). The
experimental systems used to emulate these models falls into two
distinct classes. The first involves ultracold neutral atoms in
optical lattices and the second consists of trapped ions. While the
former experimental systems allow easy realization of the BHM in
higher dimensions, the latter allow better local control on the
parameters of the model emulated \cite{cirac1}. In such an ion trap
based emulator for the BHM, phonons originating from the motional
quanta of the ions play the role of bosonic degrees of freedom. It
was pointed out in Ref.\ \onlinecite{cirac1} that the low-energy
behavior of these phonons can be described by an effective BHM. The
ground state phase diagram of such a system displays a quantum phase
transition between the Mott insulating and the superfluid phases
\cite{subir1,cirac1}. Moreover, such systems also allows us to study
the non-equilibrium dynamics of the model emulated; such dynamics
following a local quench of the on-site interaction can be observed
by looking at specific experimentally relevant observables
\cite{tarun}. The chief advantage of an ion trap emulator lies in
the fact that it allows for site-specific tuning of the on-site
interaction between the phonons; for example, this interaction can
be selectively made negative at specific sites which may lead  to
several interesting phenomena \cite{duan1,braun}. However, the
consequence of turning on such a site-specific negative local
interaction dynamically with a finite ramp rate has not been yet
studied theoretically.

In this letter, we show that turning on a site specific attractive
on-site interaction $-U_i$, where $i$ denotes the site index, with a
finite ramp time $\tau$ can generate specific entangled pure states
of the phonons {\it in a single operation}. We note that although in
this letter, we shall limit the discussions to the generation of
computationally important Bell-state between two sites, the method
is not restricted to those states only. We also show that such
dynamics involving local attractive potential provides a technique
to cool a large string of ions to their transverse motional ground
state. We contend that such cooling is a viable alternate to
resolved sideband and sympathetic cooling since it, unlike the
latter techniques, does not require the presence of multiple species
of ions and leads to a cooling time which is {\it independent of the
electronic structure of the ion species used}. We compare and
contrast our schemes of both dynamic generation of entangled states
and the proposed cooling method with the earlier ones and also
provide schematics of concrete experiments based on Barium ions in a
linear Paul trap for realization of these schemes. To the best of
our knowledge, our work constitutes the first concrete proposal of
using non-equilibrium dynamics of the BHM for both generating
computationally important pure entangled quantum many body states
and achieving ground state cooling for ions in a trap; therefore it
is expected to be of significant interest to both the
experimentalists and theorists studying viable large scale quantum
computation architecture as well as non-equilibrium dynamics of
strongly correlated systems. We note here that earlier studies of
classical dynamics in ion-trap systems are based on semi-classical
Langevin dynamics \cite{Lee}  and are significantly different from
the current work in terms of both methods used and results obtained.

\begin{figure}
\rotatebox{0}{\includegraphics*[width=0.55\linewidth]{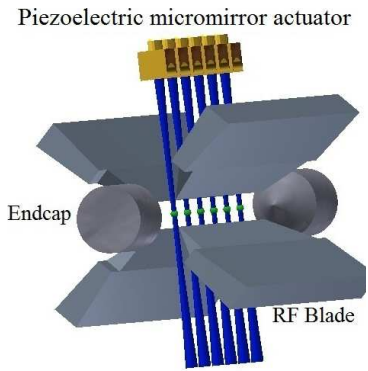}}
\caption{(Color online) A schematic of the experimental setup for
the implementation of the ramp protocol. The blue standing wave
laser phase is controlled by micro mirrors placed on peizo stages.}
\label{Fig1a}
\end{figure}

We begin by the description of a concrete experimental setup which
may serve as test bed of our proposals.  For definiteness, we
propose a linear chain of barium ions trapped and Doppler cooled in
a linear Paul trap as described in Ref.\ \cite{Steele,tarun,dubin}
as the quantum emulator for the BHM. The proposed trap is operated
at $15~$MHz radio-frequency and a trap stability parameter $q \sim
0.42$ is used for radial confinement. This generates a secular
frequency of $\omega_x \simeq 2.25~$MHz. The confinement in the
axial direction is achieved by DC voltages applied to the endcap
electrodes as shown in Fig.\ \ref{Fig1a}. This can be made shallow
so that the axial frequency is $\sim 180~$kHz leading to an
inter-ion mean distance $\sim 20~\mu$m~\cite{James} which, in turn,
leads to a tunneling strength for the transverse motional mode
phonons of $J \simeq 0.55$~kHz and $\beta_{x}=2 J/\omega_x \sim 5
\times 10^{-4}$. The inter-ionic distance in such a linear trap
varies along the chain; however, such a variation can be neglected
for  $\beta_{x}\ll1$ \cite{cirac1}. The reported heating rates in such a system is
$\omega_{\rm heat}\sim 3$ Hz \cite{heating1}; thus both the ramp
rate for phonon dynamics and the measurement cycle needs to be $\gg
\omega_{\rm heat}$ to avoid decoherence.

The barium ions after being Doppler cooled, have mean phonon number
$\bar{N}_{\rm ph} \le 10$. The state of each ion in the string are
defined by their internal ( $S$ and $D$ states for our purpose) and
external motional states (one axial and two radial states). The
external motional states are ideally decoupled from each other. In
the parameter space of interest, the radial motional mode phonons at
each site can be considered to be filling up levels of a harmonic
oscillator at individual lattice sites (defined by the ion
position). These phonons can be made to self-interact if the
oscillator is anharmonic; such a local anharmonic potential is
generated by a standing wave laser field of wavenumber $k$
interacting with the ions and leading to an on-site phonon
interaction $U = F\cos^2(k x_i)$, where $x_i$ denote ion
coordinates. Such a interaction term, along with the condition of
phonon number conservation, as described in Ref.\ \cite{tarun},
leads to $U= 2 (-1)^{\delta} F \eta_{x}^4$, where $F$ and $\delta$
are the strength of the dipole and the phase of the standing wave
formed by the laser at the ion's position respectively. The
Lamb-Dicke parameter along $x$ is denoted by $\eta_x$. Here $F$
depends on the intensity of the laser; in a typical experimental
setup, one has a $120$mW argon ion laser focussed to a $5\mu$m beam
waist on individual ions which eventually allows individual
addressing. In such a setup, it is easily possible to access the
parameter range $2 \ge J/U \ge 0.1$. Also, most importantly, $U$ can
be made repulsive or attractive by dynamically tuning the local
laser phase at each individual site \cite{James}. It is well-known
\cite{cirac1,duan1} that the effective Hamiltonian which determines
the low-energy property of the phonons is given by the BHM
\begin{eqnarray}
H &=& J \sum_{\langle ij\rangle} (b_i^{\dagger} b_j +{\rm h.c.}) +
\sum_{i} U_i {\hat n}_i({\hat n}_i -1),  \label{ham1}
\end{eqnarray}
where $b_j$ denotes the annihilation operator of the bosons
(phonons) at site $j$ and ${\hat n}_i= b_i^{\dagger} b_i$ is the
local density operator. Note that the hopping term has a positive
sign which is in contrast to the standard BHM realized with
ultracold atoms in optical lattices.

To study the dynamics, we consider a linear time evolution of a
system of $L$ Barium ions with $N$ phonons according to the protocol
$U_i(t)= U^{(0)} + (U^{(1)}-U^{(0)})t/\tau$. The time variation of
$U_i$ starts at an initial time $t=0$ with $U_i=U^{(0)}$ and
continues till $t=\tau$ when $U_i=U^{(1)}$ and is characterized by
the rate $\tau^{-1}$. The choice of sites $i$ at which the
interaction parameter is dynamically changed depends on the protocol
and shall be detailed later for specific cases. At $t=0$, we choose
a fixed $J/U^{(0)}$ (where $U^{(0)}> 0$) at each site which
corresponds to the delocalized ground state of bosons and keep the
total number of phonons(bosons) fixed to $N$. The delocalized ground
state of the bosons corresponds to having finite amplitude of boson
wavefunction on all sites. We use exact diagonalization method for
the finite-size system keeping $n \le N$ boson states per site to
obtain the energy eigenstates $|\alpha\rangle$ and eigenvalues
$E_{\alpha}$ for $H(t=\tau)$. In terms of these, one can express the
initial ground state $|\psi_G\rangle$ as $|\psi_G\rangle =
\sum_{\alpha} c_{\alpha}^0 |\alpha\rangle$, where the coefficients
$c_{\alpha}^0$ denote the overlap of the initial ground state of the
system (also obtained using exact diagonalization) with
$|\alpha\rangle$. The time-dependent Schr{\"o}dinger equation for
the system wavefunction $|\psi(t) \rangle = \sum_{\alpha}
c_{\alpha}(t) |\alpha \rangle$ governing the dynamics of the system
now reduces to equations for time evolution of $c_{\alpha}(t)$: $i
\hbar
\partial_t \sum_{\alpha} c_{\alpha}(t) |\alpha\rangle = H(t)
\sum_{\alpha} c_{\alpha}(t) |\alpha\rangle$ with the boundary
condition $c_{\alpha}(0)=c_{\alpha}^0$. To solve these equations, it
is convenient to rewrite $H(t) = H(\tau) + \Delta H(t)$ where
$\Delta H(t) = \sum_i [U_i(t)-U^{(1)}] {\hat n}_i ( {\hat n}_i -1)$.
With this choice, one obtains
\begin{eqnarray}
(i \hbar \partial_t - E_{\alpha}) c_{\alpha}(t) &=& \sum_{\beta}
\Lambda_{\alpha \beta}(t) c_{\beta} (t) \label{schtimedep}
\end{eqnarray}
where $\Lambda_{\alpha \beta}(t) = \langle \beta | \Delta H(t) |
\alpha\rangle$. The set of coupled equations for $c_{\alpha}(t)$ are
solved numerically leading to an exact numerical solution for the
time-dependent boson wavefunctions.

\begin{figure}
\rotatebox{0}{\includegraphics*[width=\linewidth]{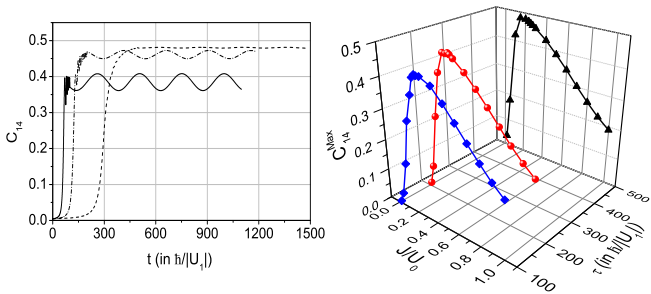}}
\caption{(Left)Plot of $C_{14}$ as a function of elapsed time and
$\tau |U^{(1)}|/\hbar= 100$ (solid), $200$ (dash-dotted), and $500$
(dashed) for $L=6$ ions with total $N=2$ phonons and
$J/U^{(0)}=0.2$. The ramp continues till $t=\tau$; see text for
details. (Right) (Color online) Plot of $C^{\rm Max}_{14}$ as a
function of $J/U^{(0)}$ for $\tau |U^{(1)}|/\hbar = 100$(blue
square), $200$(red circles) and $500$(black triangles).}
\label{Fig1}
\end{figure}

In order to generate a Bell-state of the transverse motional mode
phonons involving any two sites (say $k$ and $l$; typically chosen
to be the first and fourth sites of a linear trap of six ions with
site numbering starting from zero), we start from a fixed
$J/U^{(0)}$ and ramp the interaction on sites $k$ and $l$ to
$U^{(1)} < 0$ (chosen to be $J/U^{(1)} = -0.2$ for definiteness in
all numerics) on these sites. The interaction on other sites are
kept to $U^{(0)}$. Note that if this protocol is carried out
adiabatically with total $N$ bosons, it would lead to the Bell state
\cite{comment1}
\begin{eqnarray}
\ket{\psi_{\rm Bell}} &=&
\frac{1}{\sqrt{2}}\left(\ket{0N0000}+\ket{0000N0}\right),
\label{bells}
\end{eqnarray}
which is the ground state of the final Hamiltonian of the system. In
what follows, we carry out the ramp with a definite rate $\tau^{-1}$
and monitor the time-dependent cross correlation $C_{kl}(t) =
\langle \psi(t)|\{(b_k^{\dagger} b_l)^N+c.c\}/2|\psi(t) \rangle/N!$ between
the sites $k=1$ and $l=4$ \cite{supp1}. Since for the present case,
the non-zero value of such a correlation is equivalent to presence
of entanglement, a plot of $C_{14}$ as a function of time for
several representative ramp rates provides us with a measure of
entanglement generated at any time $t$ during or after the ramp. We
find that as the ramp is switched on, $C_{14}$ initially grows and
then saturates as shown in Fig.\ \ref {Fig1}(Left). For a fast ramp,
$C_{14}$ increases fast but saturates to a value $C_{14}^{\rm Max}$
which is much less than $0.5$ signifying that the time needed for
all the phonons to hop to the specified sites are longer than the
ramp time. Such a non-adiabatic ramp therefore cannot produce a
state which has significant overlap with $|\psi_{\rm Bell}\rangle$.
As our ultimate goal is to perform quantum gate operation using the
protocol, we search for the shortest ramp time which maximizes such
overlap leading to $C_{14}^{\rm Max} \simeq 0.5$ (the maximum for a
Bell state). To this end, we consider a system with $L=6$ ions and
$N=2$ total phonons, and vary initial value of $J/U^{(0)}$ at $t=0$
to extract the dependence of $C_{14}$ on this parameter. The results
are shown in Fig.\ \ref{Fig1}(Right) for three different
representative ramp rates. For each ramp rate, we find that
$C_{14}^{\rm Max}$ attains a maximum value for an optimal
$J/U^{(0)}$. Within the range of $\tau$ that we have studied, we
find that $C_{14}^{\rm Max} \ge 0.48$ is achieved for $J/U^{(0)}
\simeq 0.18 $ and ramp time $\tau \simeq 500 \hbar/U^{(0)}$. We
expect the presence of such an optimal $J/U^{(0)}$ to be
qualitatively unaltered for larger $L$ and $N$ for the following
reason. For $J/U^{(0)}=0$, $[H,{\hat n}_i]=0$ and the system does
not evolve due to change of $U$; thus we expect the dynamics to be
ineffective for small $J/U^{(0)}$. For $J/U^{(0)} \ge 1$, the bosons
would tend to delocalize before the system could attain the Bell
state during the dynamics. Thus we expect the dynamics to yield
optimal result for $0<J/U^{(0)}<1$ for any $L$ and $N$. We note that
a similar protocol may lead to the $W$ state where the desired state
is $\ket{\psi_W} =
\frac{1}{\sqrt{3}}\left(\ket{00N00000..}+\ket{0000N000..}+\ket{000000N0..}\right).$
Here, the protocol would involve ramping the potential to $U^{(1)}$
with $J/U^{(1)} <0$ at three chosen sites (taken to be second,
fourth and sixth sites of the chain for the state given above). A
detailed analysis of the optimal ramp rates and cross correlation
functions for such a state is left for future work.

\begin{figure}[bbb]
\rotatebox{0}{\includegraphics*[width=\linewidth]{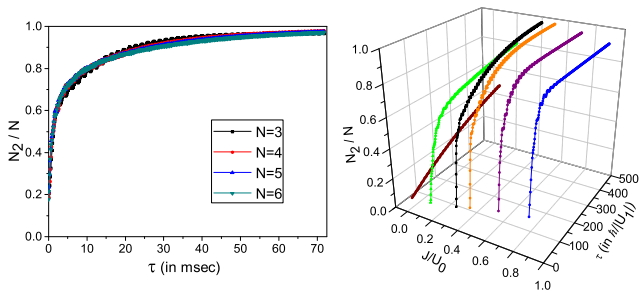}}
\caption{(Left)(Color online) Plot of $N_2(t=\tau)/N$ as a function of ramp time
$\tau$ in ms for $U^{(1)}=-1.1$ kHz,
$J/U^{(0)}=0.5$, and $N=3..6$. (Right)(Color online) Optimization of
the cooling: Plot of $N_2/N$ as a function of ramp time for
$J/U^{(0)}=0.05, 0.2, 0.3, 0.5, 0.7, {\rm and}~ 0.9$ (from left to
right) for $N=3$.} \label{Fig4}
\end{figure}

Next, we discuss a protocol for cooling. For this, we prepare a
linear chain of $L=8$ ions with $N$ transverse motional mode phonons
in superfluid state with $J/U^{(0)}\approx 0.5$ at all sites. The
protocol here involves ramping $U$ to negative values at one of the
sites (chosen to be the second site of the chain for clarity with
$J/U^{(1)}=-0.2$). This leads to migration of transverse motional
mode phonons to that site and hence to their single site confinement
leaving the rest of the chain in its motional ground state. This is
demonstrated in Fig.\ \ref{Fig4}(Left), where $N_2$
is the number of phonons on the second site at $t=\tau$, is plotted as function
of ramp time for $3\le N\le 6$. At least for low total transverse phonon
number, we find the the rate of cooling to be independent of the
total number; thus we expect our result to hold for $N\ge 6$ as
well. We also find that it takes $\sim 25 (70)$ms (with
$U^{(1)}=-1.1$kHz) for the system to have $90\% (97\%)$ overlap with
the final ground state (for which $N_2=N$). The cooling rate is a
function of both $J/U^{(0)}$ and $\tau^{-1}$; we thus optimize these
parameters to obtain the best possible cooling which is shown in
Fig.\ \ref{Fig4}(Right) for $N=3$. We note that the ramp essentially
leads to localization of excitation energies (phonons) to a single
ion site, and hence to energy reduction of other sites without
involving dissipative mechanism. This mechanism is therefore
expected to be effective for a large chain of ions a part of which
is used, for example, as a qubit since it may be used to remove
transverse motional mode energy from the computationally important
qubit states located at specific section of the chain.

The precise experimental steps for generation of the entangled Bell
state and cooling are as follows. First, $L=8$ ions of barium are
loaded into a linear ion trap forming a chain along the axis of the
linear trap. Second, these ions are then Doppler cooled to mean
phonon numbers of about $N=6$. Third, a standing wave laser at $476$
nm is formed along the transverse direction of the trap with the
ions at the anti-node. The laser power is adjusted such that the
$J/U^{(0)} \sim 0.2$ is obtained. Fourth, for the formation of one
of the Bell-states as mentioned in Eq.\ \ref{bells} between sites
$k=1$ and $l=4$, the $476$ nm standing wave laser is phase shifted
by $\pi$ phase (node) by a piezo-mounted retro-reflecting mirror.
Similar procedure is adapted for the 'W' state formation (cooling)
with $U_i$ changed for three (one) specific sites (site) as
mentioned before. In all cases, the total time of the ramp,
$\tau^{-1}$, and $J/U^{(0)}$ shall determine the fidelity of the
state obtained and the speed of the gate operation and/or cooling.
The quantities can be varied, as shown in our numerical studies
above, to obtain an optimal operating point for Bell state
generation/cooling.

The main difference of our proposals for state preparation and
cooling as compared to other proposals with trapped ions lies in its
use of a dynamic ramp. Unlike the original Cirac and Zoller's
proposal \cite{cirac2}, it is not necessary to apply sequence of
laser pulses to generate a pure many body quantum state or to
initialize the qubits. Though the M{\o}lmer and S{\o}rensen
\cite{Molmer} type of quantum gate operation does not require
initialization or ground state cooling, they require sequential
pulses to be applied in order to create a many body pure quantum
entangled state. In contrast, we do not need such elaborate
sequence. Very recently, there has been a proposal to look for
entanglement growth for 1D ultracold atom system in optical lattices
after a quench \cite{Daley}. However, such a proposal, in contrast
to ours, do not provide deterministic entangled state formation.
Regarding cooling, the most extensively used technique is the
resolved side band cooling which requires addressing of all the ions
\cite{rev1}; in contrast, the ramp protocol described here addresses
an individual site. Also, compared to cavity sideband cooling, it
does not require complicated cavity setup. Moreover, in stark
contrast to the available motional state cooling techniques, the
ramp protocol is free from use of metastable states to resolve the
motional sidebands. For such techniques, the cooling time strongly
depends on metastable state lifetime and varies between $300-30$ms
(for Hg$^+$ and Be$^+$ ions) \cite{rev1}; in contrast, our method
leads to a cooling time ($70$ ms to reach $\sim 97\%$ of the
transverse motional ground state) which is independent of the ion's
electronic structure. Thus it constitutes an alternative ground
state cooling method of a large ion string where all but one can be
used as qubits.

In conclusion, we have shown that dynamic ramp of phonons emulating
the BHM in a linear chain of trapped ions is capable of producing
computationally important maximally entangled state among large
number of qubits. A very similar protocol can also perform ground
state cooling of the transverse motional modes of a large chain of
ions. We have provided details of the dynamic ramp protocol required
for such operations and have also charted out the optimal parameter
regime for implementing them. We have shown that both these
processes can be implemented by relatively straightforward local
protocols which are well within current experimental capability \cite{haze} and
provided a comparison of our proposal to the existing ones for both
cooling and entangled state generation. We expect these protocols to
be of use in future quantum computer architecture using these
systems.

\section{Supplementary material}

In this supplementary section, we provide the details of the
correlation function $C_{kl}(t)$ between the different sites $k$ and
$l$ of a linear chain of ions and discuss how the property of such a
correlation function may be used to decipher the state at the end of
the drive protocol. We also compute the fidelity (overlap) of the
state $|\psi(t)\rangle$ with respect to Bell state and show that it
approaches unity at $t \simeq \tau$.

The correlation function $C_{kl}(t)$ between the bosons at site $k$
and $l$ of a linear chain of $L$ ions is defined in the main text,
and is given by
\begin{eqnarray}
C_{kl}(t) = \langle \psi(t)|\{(b_k^{\dagger} b_l)^N+c.c\}/2|\psi(t)
\rangle/N!,
\end{eqnarray}
where $b_k$ denotes the boson annihilation operator at site $k$ and
$N$ is the total number of bosons occupying the sites. In what
follows, we shall present numerical results for $L=6$ and $N=2$ in
accordance with the main text; however, the discussion presented
here holds for arbitrary $L$ and $N$.

\begin{figure}
\rotatebox{0}{\includegraphics*[width=\linewidth]{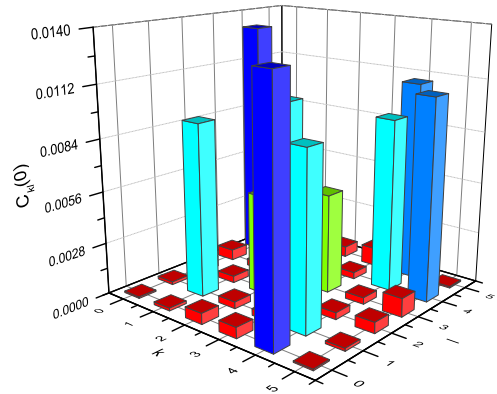}}
\caption{(Color online) Plot of the correlation $C_{kl}(0)$ for
$J/U^{(0)}=0.2$ for all sites $k$ and $l$. Here we have chosen $N=2$
and $L=6$.} \label{fig1}
\end{figure}

We begin the correlation function $C_{kl}(0)$. At $t=0$, the
interaction strength is $U_i=U^{(0)} >0$ at each site with
$J/U^{(0)} =0.2$. We have checked numerically that with these
parameters the ground state of the system, $|\psi(0)\rangle$,
represents a delocalized state of the bosons over various sites.
Such a state has finite but small amplitude $d_{n_1 n_2..n_L}$
distributed over occupation numbers $n_i \le N $ at each site $i$.
Since any state of the $N$ boson system at an arbitrary time $t$ can
be written in the occupation number basis in terms of these
coefficients as

\begin{eqnarray}
|\psi(t)\rangle = \sum_{n_i \le N} d_{n_1 n_2...n_L}(t) |n_1,n_2
...n_L \rangle, \label{state1}
\end{eqnarray}
it is easy to see that the correlation function $C_{kl}(t)$ is given
by
\begin{eqnarray}
C_{kl}(t) &=&  \Re [ \sum_{n_k \le N} d_{n_1 n_2.n_k=0
..n_L}^{\ast}(t) d_{n_1 n_2.n_l=N ..n_L}(t)]. \nonumber\\
\label{cor1}
\end{eqnarray}
Thus for $|\psi(0)\rangle$, where almost all $d_{1 2..L}^{n_1
n_2..n_L}(0)$ are finite but small, we expect a finite but small
value of $C_{kl}(0)$ for several values of $k$ and $l$. This
expectation is corroborated in Fig.\ \ref{fig1} where we find that
for $L=6$ and $N=2$, $|C_{kl}(0)| < 0.014$ for all values of $k$ and
$l$.

\begin{figure}
\rotatebox{0}{\includegraphics*[width=\linewidth]{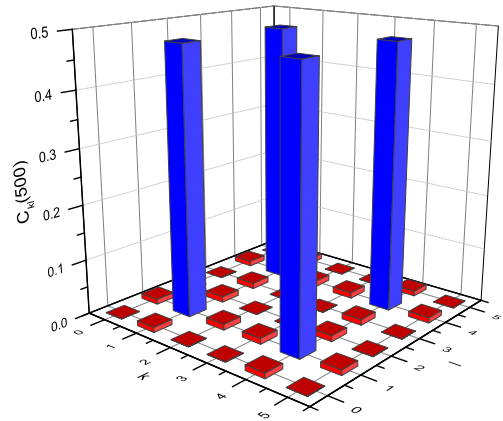}}
\caption{(Color online) Plot of the correlation $C_{kl}$ for
$J/|U^{(1)}|=0.2$ with $U^{(1)}<0$ for $k=1$ and $4$ and $U^{(1)}>0$
otherwise. For the current figure $\tau U^{(0)}/\hbar=500$,
$J/U^{(0)}=0.2$, $N=2$ and $L=6$.} \label{fig2}
\end{figure}

Next, we consider the behavior of such a correlation function for
the Bell state given by $|\psi\rangle_{\rm Bell} = [|00
n_a=N...00\rangle + |000...n_b=N..00\rangle]/\sqrt{2}$. Here the
bosons are localized either on site $a$ or $b$. Comparing
$|\psi\rangle_{\rm Bell}$ with Eq.\ \ref{state1}, we find that the
state has only two non-zero coefficients given by
$d_{00..n_a=N..00}=d_{000...n_b=N..00} = 1/\sqrt{2}$. Consequently
one finds $C_{kl}=1/2$ for $k,l=a,b$ and zero otherwise. Note that
the finite off-diagonal value of $C_{kl}$ is a consequence of the
linear superposition of the two localized bosons states $|00
n_a=N...00\rangle$ and $|000...n_b=N..00\rangle$ and hence, in this
case, a measure of the entanglement of the state. Thus the
correlation matrix $C_{kl}$ uniquely distinguishes between a
localized state with finite entanglement and delocalized boson
states.

\begin{figure}
\rotatebox{0}{\includegraphics*[width=\linewidth]{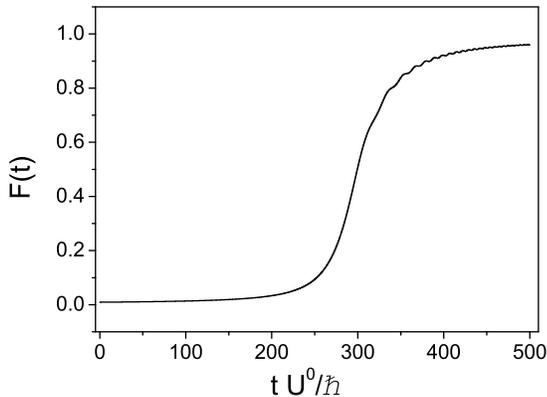}}
\caption{(Color online) Plot of $F(t)$ as a function of time $t$ for
$\tau U^{(0)}/\hbar=500$ and $J/U^{(0)}=0.2$.} \label{fig3}
\end{figure}

Next, we consider $C_{kl}^{\rm Max}$ as detailed in the text. Here
we note that since $U^{(1)}=-U^{(0)}$ on the two sites $a$ and $b$
and $U^{(1)}=U^{(0)}$ on other sites and $J/U^{(0)} \ll 1$ on all
sites, the ground state corresponding to the final Hamiltonian is
$|\psi\rangle_{\rm Bell}$. We have explicitly checked this
numerically for $L=6$ and $N=2$. Thus a knowledge of $C_{kl}^{\rm
Max}$ helps us to find the overlap of the final state after the
drive protocol with $|\psi\rangle_{\rm Bell}$. The plot of
$C_{kl}^{\rm Max}=C_{kl}(t=500U^{(0)}/\hbar) \equiv C_{kl}(500)$,
shown in Fig.\ \ref{fig2}, shows that the correlation function has
exactly the same structure as that expected from $|\psi\rangle_{\rm
Bell}$. In particular, we find that $C_{11}^{\rm Max}=C_{44}^{\rm
Max}=C_{14}^{\rm Max}=C_{41}^{\rm Max} \simeq 1/2$ and all other
$C_{kl} \simeq 0$. This shows that the state after the drive
protocol has near perfect overlap with $|\psi\rangle_{\rm Bell}$ and
thus the chosen drive protocol leads to generation of an entangled
state of bosons.

Finally, we address the fidelity of the state obtained by the ramp.
In the present context, the fidelity of the state at any time during
the dynamics is given by its overlap with the Bell state. To look at
this quantity, we define $F(t) = |\langle \psi(t)|\psi_{\rm
Bell}|\rangle|$ and study its time evolution for a specific ramp
rate $\tau U^{(0)}/\hbar=500$ and $J/U^{(0)}=0.2$. As shown in Fig.\
\ref{fig3}, we find that $F(t)$ grows and approaches unity showing a
near perfect realization Bell state for $t \simeq \tau $. The
behavior of $F(t)$ is found to be qualitatively similar for other
ramp rates and initial values of $J/U^{(0)}$.


\begin{thebibliography}{99}
\bibitem{comment1} With $|U^{(1)}|\gg J$ and $U^{(1)} <0$ on the first
and the fourth sites and positive otherwise, the ground of the
system would correspond to the bosons being localized either on site
$1$ or $4$ leading to the superposition state given by Eq.\
\ref{bells}. We have checked this explicitly using exact
diagonalization.
\end{thebibliography}
\end{document}